\documentclass{article}
\usepackage{mathrsfs}
\usepackage{graphicx}

\usepackage[final]{neurips_2025}

\usepackage[utf8]{inputenc} % allow utf-8 input
\usepackage[T1]{fontenc}    % use 8-bit T1 fonts
\usepackage{hyperref}       % hyperlinks
\usepackage{url}            % simple URL typesetting
\usepackage{booktabs}       % professional-quality tables
\usepackage{amsfonts}       % blackboard math symbols
\usepackage{nicefrac}       % compact symbols for 1/2, etc.
\usepackage{microtype}      % microtypography
\usepackage{xcolor}         % colors
\usepackage{enumitem} % for customizing lists
\usepackage{amsmath} 
\usepackage{natbib}
\setcitestyle{numbers,square}
\setlist[itemize]{noitemsep, topsep=0pt}
\usepackage{diagbox} 
\usepackage{float}
\usepackage{tablefootnote}
\usepackage{multirow}
\usepackage{multicol}
\usepackage{adjustbox}
\usepackage{subfig}
\usepackage{graphicx}
\usepackage{caption}
\usepackage{longtable}

\title{A Short Overview of Multi-Modal Wi-Fi Sensing}

\author{%
  Zijian Zhao \\
  zzhaock@connect.ust.hk \\
  Department of Civil and Environmental Engineering \\
 The Hong Kong University of Science and Technology \\
}

\begin{document}

\maketitle

\begin{abstract}
Wi-Fi sensing has emerged as a significant technology in wireless sensing and Integrated Sensing and Communication (ISAC), offering benefits such as low cost, high penetration, and enhanced privacy. Currently, it is widely utilized in various applications, including action recognition, human localization, and crowd counting. However, Wi-Fi sensing also faces challenges, such as low robustness and difficulties in data collection. Recently, there has been an increasing focus on multi-modal Wi-Fi sensing, where other modalities can act as teachers, providing ground truth or robust features for Wi-Fi sensing models to learn from, or can be directly fused with Wi-Fi for enhanced sensing capabilities. Although these methods have demonstrated promising results and substantial value in practical applications, there is a lack of comprehensive surveys reviewing them. To address this gap, this paper reviews the multi-modal Wi-Fi sensing literature \textbf{from the past 24 months} and highlights the current limitations, challenges and future directions in this field. \footnote{In this paper, we only focus on the learning-based methods but ignore those model-based ones. As Wi-Fi sensing is not as popular as CV and NLP, plus the multi-modal Wi-Fi sensing is just a small branch in it, the works we reviewed in this short survey comprise only about 30.} 
\end{abstract}

\section{Introduction}

In recent years, Wi-Fi sensing has emerged as a transformative technology within the broader field of Integrated Sensing and Communications (ISAC), offering a novel approach to enhance the efficiency and intelligence of communication systems. At its core, Wi-Fi sensing leverages the existing infrastructure of Wi-Fi networks to perform sensing tasks such as motion detection, gesture recognition, biometric measurement, and environmental monitoring. By analyzing variations in Wi-Fi signals—such as changes in received signal strength (RSSI) and channel state information (CSI)—caused by interactions with objects and the environment, Wi-Fi sensing enables a wide range of applications without the need for additional dedicated sensors. This technology is particularly appealing due to its contactless nature, privacy-preserving capabilities, insensitivity to occlusions and lighting conditions, ubiquity, and cost-effectiveness compared to specialized radar systems.

\begin{figure}[htbp]
\centering 
\includegraphics[width=0.7\textwidth]{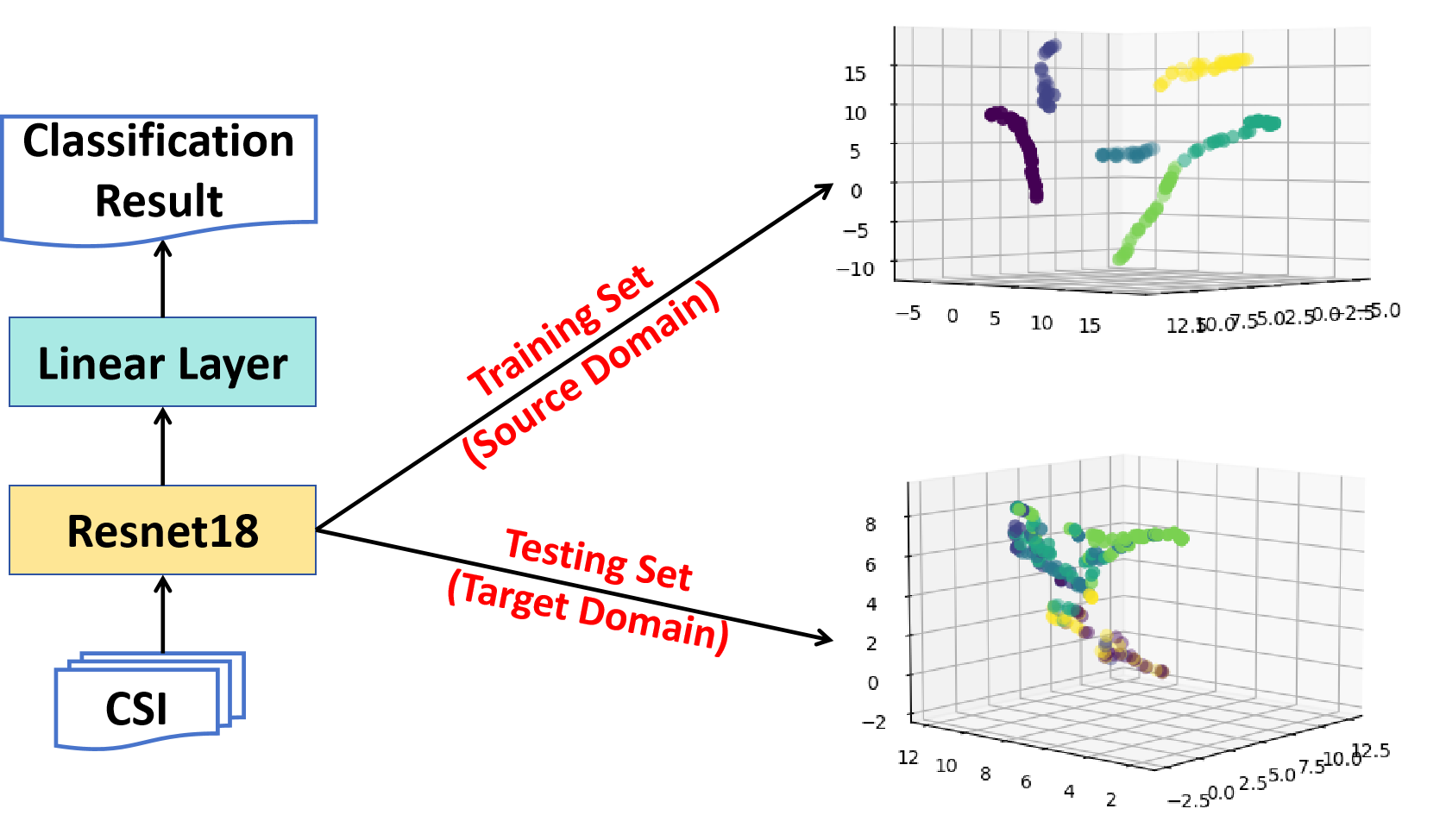}
\caption{Cross Domain Challenge Visualization in \cite{zhao2024knn}: Different colors represent different categories.}
\label{fig:cross-domain}
\end{figure}

The potential of Wi-Fi sensing extends across diverse domains, including elderly care (e.g., fall detection) \cite{chen2024deep}, security (e.g., intrusion detection) \cite{ou2025codar}, smart environments (e.g., person identification) \cite{cao2025real}, and healthcare (e.g., breath detection) \cite{fan2024contactless}. Its ability to operate through walls and in low-light conditions further enhances its applicability in scenarios where traditional sensing methods, such as cameras, may be limited. However, despite its promise, Wi-Fi sensing faces significant challenges, particularly in terms of robustness. Existing models often exhibit limited generalizability, with even minor environmental changes—such as alterations in room layout or the presence of new objects—leading to substantial performance degradation or complete failure. This challenge is especially serious for learning-based methods. An example is shown as Fig. \ref{fig:cross-domain} \cite{zhao2024knn}, where the network trained in the source domain can hardly extract features and distinguish them in the target domain. Addressing this robustness issue is critical for the practical deployment of Wi-Fi sensing systems.

Recently, many researchers have shifted towards multi-modal Wi-Fi sensing, aiming to let the Wi-Fi modality learn from other strong modalities, such as radar, LiDAR, and vision-based systems, or to directly use them together to combine their benefits. By fusing data from multiple sources, multi-modal Wi-Fi sensing aims to enhance the accuracy, reliability, and adaptability of sensing tasks. For instance, combining Wi-Fi CSI with radar data can provide complementary information, enabling more robust and detailed environmental understanding \cite{chen2025research}. This multi-modal approach not only mitigates the impact of environmental variability but also opens up new possibilities for complex applications.

However, we noticed there is currently no review focusing on the field of multi-modal Wi-Fi sensing. To fulfill this gap, this short review provides a comprehensive overview of the latest 24 months of related works, especially focusing on multi-modal deep Wi-Fi sensing. Finally, we provide a perspective discussion about the current limitations, challenges, and future directions. By synthesizing the state-of-the-art in this rapidly evolving field, we aim to provide a foundation for future innovations that will drive the practical deployment of Wi-Fi sensing in real-world scenarios. Additionally, in this paper, we do not include the multi-modal datasets part since it has been summarized in the latest survey about Wi-Fi sensing generalizability \cite{wang2025survey}, which interested readers can refer to.

\section{Background of Wi-Fi Sensing}

\begin{figure}[htbp]
\centering 
\includegraphics[width=0.7\textwidth]{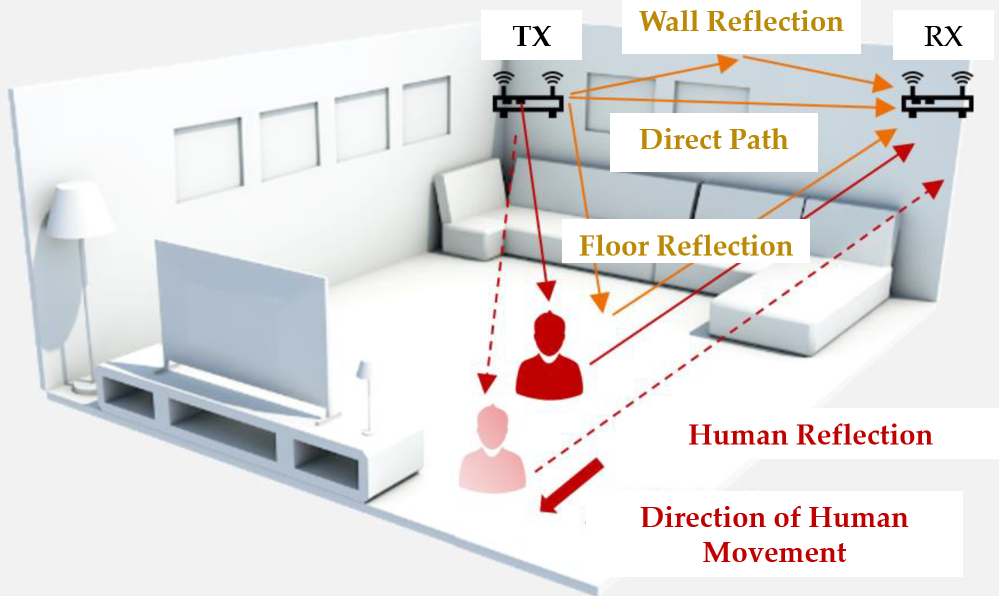}
\caption{Basic Principles of Wi-Fi Sensing}
\label{fig:principle}
\end{figure}

In Wi-Fi sensing, the two most commonly utilized features are Channel State Information (CSI) and Received Signal Strength Information (RSSI). When individuals move within the environment, the signal characteristics in the air undergo corresponding changes, which we can capture for specific sensing tasks such as action recognition and people localization, as shown in Fig. \ref{fig:principle}.

CSI is employed to provide feedback on the characteristics of a wireless channel. In scenarios where both the transmitter and receiver are located within the same indoor environment, the transmitted signal traverses multiple paths, experiencing reflections, refractions, or scattering before reaching the receiver. Mathematically, this channel can be represented as:
\begin{equation}
Y = HX + N \ ,
\end{equation}
where $Y$ and $X$ are the matrices of the received and transmitted signals, respectively; $N$ is the vector of noise signals; and $H$ is the estimate of the Channel Frequency Response (CFR) of the wireless channel, which contains information about the amplitudes, phases, and delays of the multipath components. It can also be represented as:
\begin{equation}
H=||H||e^{j \angle H} \ ,
\end{equation}
where $||H||$ and $\angle H$ denote the amplitude and phase of the CSI measurement, respectively. Additionally, the CFR $H$ can be also expressed as:
\begin{equation}
    H(f, t) = H_s(f, t) + H_d(f, t) \ ,
\end{equation}
where $f$ is the subcarrier frequency and $t$ is the time-domain sampling point. This equation can be divided into $H_s(f, t)$, the static component corresponding to the environment, and $H_d(f, t)$, the dynamic component corresponding to people's actions.

RSSI measures the power level of the received signal at the receiver, typically expressed in decibel-milliwatts (dBm). Unlike CSI, which captures fine-grained frequency-selective channel characteristics, RSSI provides a scalar value representing the aggregate signal strength across the entire channel bandwidth. In a multipath propagation environment, the received signal power fluctuates due to constructive and destructive interference among multiple signal paths. Mathematically, the RSSI at time $t$ can be modeled as:
\begin{equation}
\text{RSSI}(t) = P_t - L(t) + \eta(t) \ ,
\end{equation}
where $P_t$ is the transmitted power, $L(t)$ denotes the total path loss, and $\eta(t)$ represents noise and fast-fading effects. For narrowband systems, RSSI can also be approximated using the log-distance path loss model:
\begin{equation}
\text{RSSI}(d, t) = \text{RSSI}_0 - 10n \log_{10} \left( \frac{d(t)}{d_0} \right) + \chi(t) \ ,
\end{equation}
where $\text{RSSI}_0$ is the reference power at distance $d_0$, $n$ is the path loss exponent, $d(t)$ is the time-varying distance between transmitter and receiver, and $\chi(t)$ accounts for shadowing and multipath effects.

In general, while RSSI lacks the granularity of CSI in capturing frequency-selective fading, it remains computationally efficient and requires minimal hardware support, making it suitable for resource-constrained sensing tasks. However, its accuracy is highly susceptible to environmental noise and temporal variations. Conversely, CSI can capture more detailed information and is more appropriate for fine-grained scenarios. Although estimating CSI is more complex and resource-intensive, deep learning-based CSI prediction methods \cite{zhao2024mining} have emerged as a promising solution. Furthermore, the upcoming Wi-Fi 8 \cite{liu2024wi} may provide enhanced support for CSI capture to facilitate these sensing tasks.

\section{Main Paradigms of Multi-Modal Wi-Fi Sensing}

For current multi-modal Wi-Fi sensing works, we mainly divide them into two types. The first type is multi-modal fusion Wi-Fi sensing, where different sensors are fused together to enhance the sensing capacity or generalization capacity. The other type is multi-modal enhanced training of the Wi-Fi modality, where models from strong modalities serve as teachers or label generators, providing ground truth or dark knowledge for Wi-Fi sensing models to learn.

\begin{figure}[htbp]
\centering 
\includegraphics[width=\textwidth]{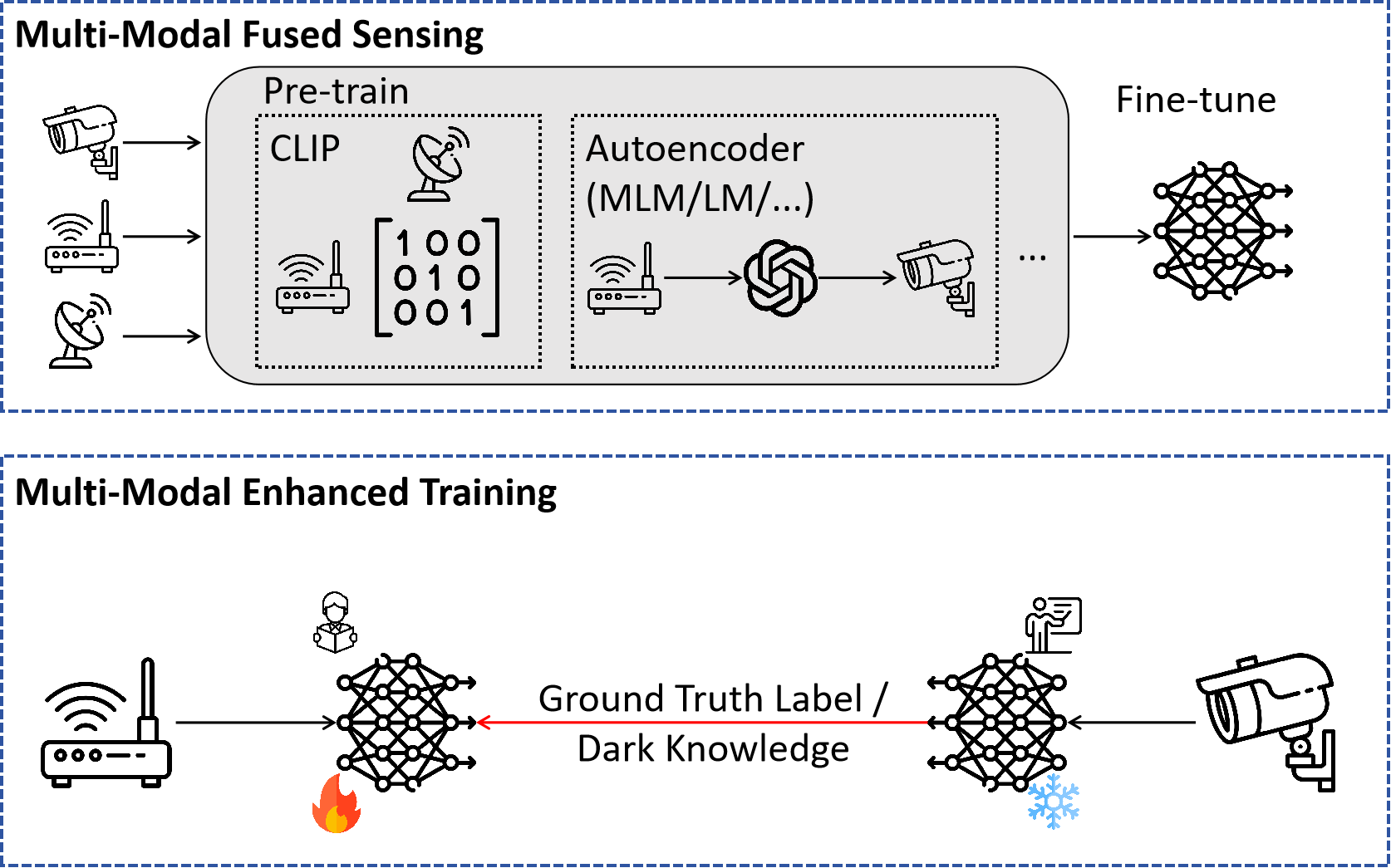}
\caption{Main Paradigms of Multi-Modal Wi-Fi Sensing}
\label{fig:main}
\end{figure}

\subsection{Multi-Modal Fused Sensing} \label{sec:Multi-Modal Fused Sensing}

As in many multi-modal fields, the multi-modal fusion Wi-Fi sensing methods can also be divided into input-fusion and feature-fusion types (we have not found any decision-level fused methods in Wi-Fi sensing). The methods are summarized in Table \ref{tab:methods}.

\begin{table}[htbp]
    \caption{Summary of Multi-Modal Fused Sensing Methods: For the pre-training methods, we only consider cross-modal pre-training and ignore single-modal encoder pre-training. The `V' represents vision, `R' represents RaDar, and `X' represents multiple sensors. The `HAR' represents human action recognition, `CC' represents crowd counting, `GR' represents gesture recognition, and `PR' represents person re-identification.}
    \centering
    \begin{adjustbox}{width=1.00\textwidth}
    \begin{tabular}{l@{\quad}|cccccccc}
        \toprule
        \textbf{Method} & \textbf{Modality} & \textbf{Wi-Fi} & \textbf{Task} & \textbf{Backbone} & \textbf{Fusion} & \textbf{Pre-training} & \textbf{Modal Alignment}& \textbf{Cross Domain} \\
        \midrule
        \textbf{SCL \cite{liu2024wireless}} & X & CSI & Multiple & GNN & Input & $\times$ & $\times$ & $\times$ \\
        \textbf{Ye et al. \cite{ye2025multi}} & V & CSI & Localization & ViT & Input & $\times$ & $\times$ & $\times$ \\
       \textbf{MaskFi \cite{yang2024maskfi}} & V & CSI & HAR & ViT & Input & MLM & $\times$ & $\times$ \\ 
       \midrule
        \textbf{Zhou et al. \cite{zhou2024heterogeneous}} & V & RSSI & Localization & CNN+LSTM & Feature & $\times$ & $\times$ & $\times$ \\
        \textbf{HDANet \cite{hao2024heterogeneous}} & V & CSI & CC & CNN & Feature & $\times$ & $\times$ &  $\times$ \\
        \textbf{ViFi-ReID \cite{mao2024vifi}} & V & CSI & PR & Transformer & Feature & $\times$ & CLIP & one-shot \\
        \textbf{X-Fi \cite{chen2024x}} & X & CSI & HAR & Transformer & Feature & $\times$ & $\times$ & $\times$ \\
        \textbf{Babel \cite{dai2025babel}} & X & CSI & HAR & Prototype & Feature & CLIP & $\checkmark$ & one-shot \\
        \textbf{Wivi-Uf \cite{lin5145405wivi}} & V & CSI & HAR & Transformer & Feature & $\times$ & $\times$ & $\times$ \\
        \textbf{WiMix \cite{chen2023wimix}} & V & CSI & HAR & CNN & Feature & $\times$ & $\times$ & $\times$ \\
        \textbf{WiFitness \cite{wei2024wi}} & V & CSI & HAR & GCN+CNN & Feature & $\times$ & SSA & $\times$ \\
       \textbf{Chen et al. \cite{chen2025research}} & R & CSI & HAR & GCN+CNN & Feature & $\times$ & $\times$ & zero-shot \\
        \bottomrule
    \end{tabular}
    \end{adjustbox}
    \label{tab:methods}
\end{table}

\subsubsection{Input Fusion Methods}

For the input fusion methods, the original data of different modalities are first combined together and then input to a single neural network. Liu et al. \cite{liu2024wireless} proposed a powerful method named SCL, which can process sensors of different modalities like RFID, Bluetooth, Wi-Fi, and Zigbee, and can be used in multiple tasks such as action recognition, person localization, and anomaly detection, serving as a general multi-modal Wi-Fi sensing framework. In detail, it first uses cross-modal attention to extract feature embeddings of each modality input and constructs a graph based on this embedding, which is then input to a GNN for specific task outputs.

In Ye et al. \cite{ye2025multi} and MaskFi \cite{yang2024maskfi}, the ViT is used to process the concatenated image and Wi-Fi CSI for localization and action recognition tasks, respectively, where the CSI is viewed as a particular type of image and can be divided into patches like an image. In MaskFi, MLM \cite{devlin2019bert} is also used for pre-training, which takes full advantage of the unlabeled data. However, the current version of the paper does not provide performance comparisons with and without the pre-training.

\subsubsection{Feature Fusion Methods}

Compared to the input-fusion method, there are more works selecting the feature-fusion approach. Due to the heterogeneous data formats of different modalities, it is challenging to design a network that can take them as input together. On the contrary, it is easier to use separate encoders to map each modality to a feature vector for subsequent processing.

For the localization task, Zhou et al. \cite{zhou2024heterogeneous} proposed using a shared encoder to process images and Wi-Fi RSSI, similar to a Siamese network \cite{bromley1993signature}, improving parameter efficiency. Hao et al. \cite{hao2024heterogeneous} introduced a novel network named HDANet for crowd counting based on Wi-Fi CSI and images, where LeNet \cite{lecun1998gradient} and ResNet \cite{he2016deep} are used to extract features from Wi-Fi and images, respectively. Self-attention and cross-attention are employed to capture relationships within single modalities and across modalities in turn. Mao et al. \cite{mao2024vifi} proposed a vision+Wi-Fi person re-identification method, where, in addition to the conventional contrastive learning loss, a CLIP loss \cite{radford2021learning} between the feature embeddings of Wi-Fi and vision is added to enhance feature representation capacity through modal alignment. (Here, the CLIP is not used as a pre-training manner, but used together during training for enhancement.)

Additionally, most works focus on the multi-modal human action recognition task. X-Fi and Babel \cite{dai2025babel} both propose novel network structures that can combine multiple different sensors like Wi-Fi, mmWave, IMU, LiDAR, images, and depth together simultaneously, presented at ICLR 2025 and SenSys 2025, respectively. In X-Fi, pre-trained feature extractors are first trained using each single modality's data separately. The extracted embeddings are then input to a cross-modality Transformer \cite{vaswani2017attention} for modality-invariant feature extraction. In each training iteration, the used modalities are randomized, allowing X-Fi to function with arbitrary missing modalities. For Babel, modality alignment is explicitly realized by CLIP between different modalities. The modality feature extractor is also pre-trained on single-modal data, and the concept alignment modules are trained for each modality using a shared prototype network via CLIP. Their method demonstrates high expandability and alignment performance. Such aligned methods help extract more common and general features across different modalities, which reduces the risk of overfitting and shows promising performance in cross-dataset (domain) scenarios under the one-shot fine-tuning setting.

In Wivi-Uf \cite{lin5145405wivi}, WiMix \cite{chen2023wimix}, and WiFitness \cite{wei2024wi}, all developed the human activity recognition framework using the extracted feature fusion from image and Wi-Fi CSI. Wivi-Uf \cite{lin5145405wivi} and WiMix \cite{chen2023wimix} both used the cross-modal attention to capture the cross modality relationship. While, WiFitness \cite{wei2024wi} use the local self-attention to extract the inner modality relationship. Additionally, it also propose a novel alignment method, named Spatio-Temporal Semantic Alignment (SSA), by minimizing the norm of correlation matrix between modalities. At last, Chen et al. \cite{chen2025research} proposed an activity recognition method based on Wi-Fi and RaDar, where the DANN method \cite{ganin2016domain} is combined to extract domain-invariant feature, realizing the zero-shot cross sense capacity.

\subsubsection{Others}
Lastly, there is a special type of multi-modal Wi-Fi sensing model that utilizes multiple different signal features derived from the same original signal. Gao et al. \cite{gao2024autosen} proposed a novel pre-training method where an autoencoder \cite{hinton2006reducing} uses CSI amplitude as input and CSI phase as the recovery target, enhancing model performance in few-shot cross-domain scenarios. Similarly, Chen et al. \cite{chen2024aiot} applied Discrete Wavelet Transform (DWT) and Short-Time Fourier Transform (STFT) to processed CSI for fine activity recognition, introducing a novel fusion branch to efficiently extract features from each transformation. (Although there are actually many similar works like \cite{zhang2023ratiofi}, we only conclude those that explicitly identify themselves as multi-modal. Additionally, there is no clear definition of whether these works can be viewed as true multi-modal.)

\subsection{Multi-Modal Enhanced Training}

For the multi-modal enhanced training method, the amount of work is relatively small, summarized in Table \ref{tab:methods2}. These methods can be mainly divided into two types: Cross-Modal Knowledge Distillation (CMKD), where the strong modalities serve as teachers for Wi-Fi, and methods where strong modalities directly generate ground truth labels for Wi-Fi sensing models.

\begin{table}[htbp]
    \caption{Summary of Multi-Modal Enhanced Training Methods}
    \centering
    \begin{adjustbox}{width=1.00\textwidth}
    \begin{tabular}{l@{\quad}|cccccc}
        \toprule
        \textbf{Method} & \textbf{Teacher} & \textbf{Wi-Fi} & \textbf{Task} & \textbf{Backbone} & \textbf{Type} & \textbf{Cross Domain}\\
        \midrule
        \textbf{Rizk et al. \cite{rizk2024precise}} & RTT & RSSI & Localization & Transformer & KD & $\times$\\
        \textbf{XFall \cite{chi2024xfall}} & Vision & CSI & Fall Detection & Transformer & KD & zero-shot \\
        \textbf{MuAt-Va \cite{sheng2023muat}} & Vision & CSI & HAR &  CNN & KD & $\times$ \\
        \textbf{AutoDLAR \cite{lu2024autodlar}} & Vision & CSI & HAR & CNN & KD & $\times$ \\
        \textbf{FallDewideo \cite{cai2023falldewideo}} & Vision & CSI & Fall Detection & CNN & Label Generator & $\times$ \\
        \textbf{LoFi \cite{zhao2024lofi}} & Vision & CSI & Localization & -- &  Label Generator & $\times$ \\
        \bottomrule
    \end{tabular}
    \end{adjustbox}
    \label{tab:methods2}
\end{table}

\subsubsection{Cross-Modal Knowledge Distillation}

Wi-Fi is a weak modality, highly dependent on the environment, which leads to low robustness and generalization capacity in Wi-Fi sensing models. Consequently, many works have begun to explore using stronger modalities as teachers for the Wi-Fi modality, with Cross-Modal Knowledge Distillation (CMKD) as a representative approach.

In classification tasks, important information is often excluded from the ground truth labels. For instance, there are varying similarities between different labels that should be reflected in the classification probabilities. Additionally, noisy samples may resemble other specific categories. If this information can be captured during training, we can enhance model performance, which is the core idea behind Knowledge Distillation (KD) \cite{hinton2015distilling}.

In standard KD, a teacher network is first trained to generate the classification probabilities for each training sample, represented as $P(\cdot|x;\theta_t)$, where $x$ is the input sample and $\theta_t$ are the parameters of the teacher network. The information contained in $P(\cdot|x;\theta_t)$ is referred to as dark knowledge, reflecting the similarities among samples across categories. This dark knowledge serves as a soft target for the student network, formulated as:
\begin{equation}
L_s^{cls} = \nabla_{\theta_s} \mathbf{E}_x \left[ \sum_{y \in Y} P_t(y|x;\theta_t) \log P_s(y|x;\theta_s) \right] \ ,
\label{eq:kd}
\end{equation}
where $Y$ is the label set, $P_s$ and $\theta_s$ are the probabilities and parameters of the student network, respectively, and $L_s^{cls}$ is called the soft loss. The soft loss is typically used alongside the target loss, which is the standard cross-entropy loss that allows the student network to learn directly from the ground truth. This approach enables the student to balance learning true knowledge from the labels and dark knowledge from the teacher network. When expanding this application to regression tasks, the soft loss can also be replaced by:
\begin{equation}
L_s^{mse} = \nabla_{\theta_s} \mathbf{E}_x \left[ ||\text{Emb}(x;\theta_t) - \text{Emb}(x;\theta_s)||_2 \right] \ ,
\end{equation}
where $\text{Emb}(\cdot)$ represents the hidden layer output of the network.

Based on the standard KD, Rizk et al. \cite{rizk2024precise} utilized a Round-Trip Time (RTT)-based localization method as the teacher network for an RSSI-based Wi-Fi positioning model, reducing the localization error by over 75\%. MuAt-Va \cite{sheng2023muat}, XFall \cite{chi2024xfall}, and AutoDLAR \cite{lu2024autodlar} employed similar methods to enable Wi-Fi sensing models to learn from videos. MuAt-Va directly used standard KD for action recognition task. XFall, used for fall detection (a binary classification task), selected the MSE format $L_s^{mse}$ for the soft loss and demonstrated excellent performance in zero-shot cross-domain scenarios. In contrast, AutoDLAR, applied to action recognition (classification tasks), combined both formats of soft loss: $L_s^{cls} + L_s^{mse}$. Notably, it directly utilized a trained computer vision model as the teacher and provided no ground truth to the student model, meaning there was no hard loss. This model showed promising results on unlabeled datasets, aligning with the methods discussed in the next subsection, where strong modalities are used to provide learning ground truth directly.

\subsubsection{Label Generation from Strong Modality}

\begin{figure}[htbp]
\centering 
\includegraphics[width=\textwidth]{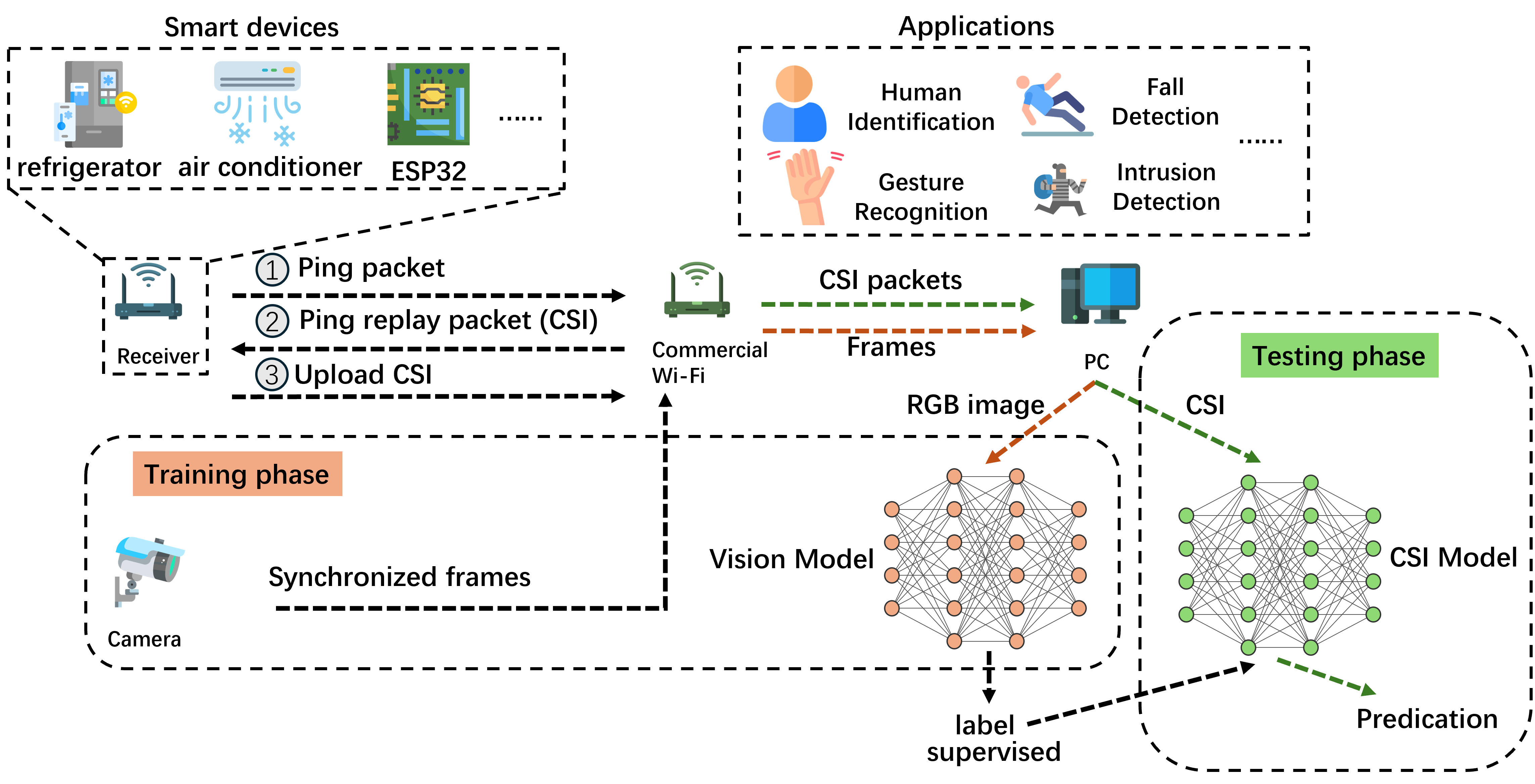}
\caption{Label Generation from Vision Modality \cite{zhao2025crossfi}}
\label{fig:vision}
\end{figure}

The challenge of limited labeled data in Wi-Fi sensing is widely acknowledged, making few-shot learning and unsupervised pre-training methods extremely popular in this field. To fundamentally address this problem, we require powerful models that can efficiently utilize heterogeneous datasets or reduce the complexity and cost of data collection. The former approach is more challenging and still seems a long way from realization. In contrast, the latter is more practical at the current stage. One efficient method is to use strong, mature modalities to generate ground truth labels for Wi-Fi sensing models. For instance, in tasks like Wi-Fi imaging \cite{xu2024wicamera}, where images serve as natural labels, extracting labels from images has proven feasible, thanks to advancements in computer vision (CV) models. A common workflow is shown as Fig. \ref{fig:vision}.

In FallDewideo \cite{cai2023falldewideo}, a workflow for a fall detection Wi-Fi sensing system was developed. Conventional methods for collecting action recognition datasets typically require a person to control the device for recording while another executes the action. This process is time-consuming, as it necessitates manual labeling and setup each time a new action is collected. In contrast, FallDewideo employs an automated system that utilizes a camera and a Wi-Fi receiver to collect data simultaneously, using a synchronous ring to indicate when data is being recorded. Additionally, a fall detection vision model \cite{hazelhoff2008video} generates labels based on the recorded video. This system eliminates the need for manual control or pauses between recording multiple samples, and it has potential applications in other action recognition scenarios, such as gesture recognition. Furthermore, FallDewideo proposes a benchmark fall detection method that includes a novel Human Pose Estimation (HPE) pre-training technique to enhance model performance in downstream fall detection tasks. Specifically, they use a Mask R-CNN \cite{he2017mask} enhanced OpenPose \cite{cao2017realtime} model to generate ground truth for joint heat maps (JHM) and part affinity fields (PAF), which are used by a U-Net \cite{ronneberger2015u} based Wi-Fi sensing model.

In our previous work, LoFi \cite{zhao2024lofi}, we proposed a vision-aided dataset collection method for Wi-Fi localization and tracking. Previously, collecting such datasets was difficult and costly. To obtain high-precision datasets, devices like LiDAR were needed, which are very expensive. An alternative method involved setting up specific points or routes in advance \cite{zhang2025leveraging}, allowing people to stand or walk on them. However, this approach limited the amount of ground truth data, hindering model training and development. Additionally, most studies focused only on coarse-grained localization tasks \cite{liu2024sifi}, where rooms are divided into zones, and models classify which zone a person is located in. While this reduces data collection difficulty, the precision often fails to meet practical application requirements. To address these issues, we proposed a vision-aided label generation method that simultaneously records Wi-Fi signals and video. We then use object detection models like YOLO \cite{redmon2016you} to determine the coordinates of individuals in pixel space. Finally, we convert pixel coordinates to real-world physical space to obtain ground truth localization. We observed that this method can achieve an error of less than 20 cm without requiring specific camera angles.

As these two methods have shown promising results and are straightforward to implement, we believe that more similar methods can be developed for related tasks, such as people identification and crowd counting.

\subsection{Mixture Methods}

At last, there is a particular type of worker combining the above two methods, where a strong modality not only serves as a teacher but is also used with Wi-Fi for fused sensing, combining the advantages of these two methods together \cite{tang2023hybrid,tang2024novel,hori2024wi}. Tang et al. \cite{tang2023hybrid,tang2024novel} proposed a localization method that uses both vision and Wi-Fi, where, in addition to the MSE loss in conventional localization tasks, a soft loss as shown in Eq. \ref{eq:kd} is added. For the student part, the multi-modal network uses another classification head for coarse-grained localization (to identify which zones the person is in). For the teacher part, a classification head is directly used with the output from the vision branch for prediction, without using Wi-Fi. In \cite{hori2024wi}, a novel Wi-Fi-based caption method is proposed, where some indirect sensors, including Wi-Fi, depth, and thermal sensors, are used together to learn from a teacher network based on direct sensors, including vision and audio. This method achieved a 28\% improvement over the Wi-Fi-only method.

\section{Discussions}

\subsection{Limitations}

Shown in Section \ref{sec:Multi-Modal Fused Sensing}, current multi-modal fused sensing methods can be divided into early fusion (input fusion) and intermediate fusion (feature fusion). Even in the general multi-modal machine learning field, there is no definitive conclusion as to which approach is better. In multi-modal Wi-Fi sensing, early fusion methods often use different data processing techniques to transform various modalities into a common format, such as constructing a common graph \cite{liu2024wireless} or viewing Wi-Fi as a specific image \cite{ye2025multi,yang2024maskfi}. This approach allows different modal data to be combined and input into a single model, fully utilizing the original data and avoiding potential information loss.

For intermediate fusion, we first extract features from each modality, respectively, and then combine them. Many works utilize cross-modal attention to capture the relationships between modalities:

\begin{equation}
    \text{CrossAttention}(Q^A, K^B, V^B) = \text{softmax}\left( \frac{Q^A (K^B)^\top}{\sqrt{d_k}} \right) V^B \ ,
\end{equation}

where

\begin{itemize}[left=0pt]
    \item $\mathbf{Q}^A = \mathbf{X}^A \mathbf{W}_Q^A \in \mathbb{R}^{n \times d_k}$: Query matrix from modality $A$
    \item $\mathbf{K}^B = \mathbf{X}^B \mathbf{W}_K^B \in \mathbb{R}^{m \times d_k}$: Key matrix from modality $B$
    \item $\mathbf{V}^B = \mathbf{X}^B \mathbf{W}_V^B \in \mathbb{R}^{m \times d_v}$: Value matrix from modality $B$
    \item $\mathbf{X}^A \in \mathbb{R}^{n \times d_A}$: Input features from modality $A$ with sequence length $n$
    \item $\mathbf{X}^B \in \mathbb{R}^{m \times d_B}$: Input features from modality $B$ with sequence length $m$
    \item $d_k$: Dimension of key vectors (scaling factor)
    \item $\mathbf{W}_Q^A \in \mathbb{R}^{d_A \times d_k}$, $\mathbf{W}_K^B \in \mathbb{R}^{d_B \times d_k}$, $\mathbf{W}_V^B \in \mathbb{R}^{d_B \times d_v}$: Learnable projection matrices
\end{itemize}

In many works, the proposed methods are only compared with conventional single-modal methods, so we currently do not know the performance difference between the two types of methods in sensing tasks. However, it deserves exploration to combine them, which can be viewed as based on the ViT of \cite{ye2025multi,yang2024maskfi}, explicitly adding cross-modal attention. Additionally, the late fusion (decision fusion) method also deserves consideration, or combining it with the other two types, since currently there seems to be no such work and we do not know whether this paradigm is useful.

Moreover, modal alignment has shown promising performance in many works \cite{mao2024vifi,dai2025babel,wei2024wi}, where CLIP has been widely used as follows:

\begin{equation}
    \mathcal{L}_{\text{A2B}} = -\frac{1}{N} \sum_{i=1}^N \log \frac{\exp(s(\mathbf{X}^A_i, \mathbf{X}^B_i))}{\sum_{j=1}^N \exp(s(\mathbf{X}^A_i, \mathbf{X}^B_j))} \ ,
\end{equation}

where $N$ is the sample amount, and $s$ represents the similarity calculation method, with cosine similarity being widely used. However, the success of this method remains unclear. Is it because modal alignment helps the encoders of each modality learn more general features? Or is it merely the case that adding such a loss function or pre-training helps modal training become more effective in this limited training set field? Further experiments comparing the modal alignment method with other pre-training methods, such as MLM, should be conducted. Combining them may also provide different insights.

Furthermore, the specific application scenarios should be clearly understood in multi-modal Wi-Fi sensing. Some modalities can undermine the advantages of Wi-Fi itself; for instance, cameras may negate the high privacy benefits of Wi-Fi, and expensive sensors can make the low cost of Wi-Fi meaningless.

For the multi-modal enhanced methods, the generalization problem requires more attention. In CMKD methods, the dark knowledge from strong modalities helps improve the generalization capacity of Wi-Fi sensing models. However, in practice, we observe that these methods also suffer from overfitting. This occurs when Wi-Fi sensing models become too aligned with the dark knowledge of the teacher modality in the source domain (training set), resulting in poor generalization to the target domain. In such cases, KD does not benefit the Wi-Fi sensing models. This dilemma is particularly prevalent when the data size is limited, which has been a longstanding issue in Wi-Fi sensing. As summarized in Tables \ref{tab:methods} and \ref{tab:methods2}, even though multi-modal Wi-Fi sensing methods have huge potential for cross-domain tasks, they have not received full discussion and attention in many works. Even with the enhancement of Wi-Fi sensing accuracy through multi-modal technology, the improvement seems somewhat limited. In in-domain scenarios, single-modal Wi-Fi sensing methods already achieve relatively high accuracy. In practice, improving accuracy from 90\% to 95\% may not have significant practical implications. For practical applications, the training and evaluation scenarios are often different, so more effort should be directed toward developing cross-domain or transfer learning methods, where the trained model can quickly adapt to a novel environment. As multi-modal methods utilize stronger modalities, they offer more benefits in such domain adaptation or generalization tasks.

What's more, an important challenge that has not been widely discussed in previous works is time and space alignment. Even for most works, the samples used are all time sequences, where small errors between modalities are acceptable. However, for future fine-grained tasks, we may need finely aligned data pairs (e.g., aligned in frames). As a result, it deserves exploration to assess this influence and develop high-precision dataset collection methods

Finally, a notable shortcoming in the Wi-Fi sensing community is that most papers do not open-source their code, and some research collects new datasets but fails to release them. While this issue is common in various fields outside pure AI—such as transportation, robotics, and communication—it is detrimental to the development of the field. We also observe that many studies struggle with reproducibility; even well-cited works in top journals and conferences can yield suboptimal results when independently replicated by other researchers. Additionally, due to the varying formats of devices, certain methods may perform well only on the datasets used in their original studies, making them difficult to transfer to other scenarios. These factors complicate the ability of researchers to identify solid work to build upon, collectively hindering progress in the field.

\subsection{Challenges}

In Wi-Fi sensing, one of the biggest challenges is the limited and heterogeneous public datasets. The different formats of data collected by various devices make it difficult to combine them for training a single model. Considering the high difficulty and consumption involved in constructing multi-modal datasets, alongside the privacy concerns for modalities like image and audio, there are fewer public datasets in the field of multi-modal Wi-Fi sensing. Pre-training and fine-tuning appear to be promising paradigms where models can learn from unlabeled datasets in an unsupervised manner. Collecting unlabeled data can be much easier and cheaper than obtaining labeled datasets. As summarized in \cite{xu2025evaluating}, current pre-training methods can be mainly divided into four types: cluster discrimination, instance discrimination, relation prediction, and autoencoder (including LM/MLM). Due to the success of Transformers in large language models (LLMs), an increasing number of studies have applied these techniques in large multi-modal models (LMMs), including in the Wi-Fi sensing field. A significant factor contributing to the success of current LLMs is the use of language model (LM) \cite{radford2018improving} and masked language model (MLM) \cite{devlin2019bert} pre-training. Since 2021, several works have introduced these approaches to Wi-Fi sensing. However, there is still no definitive conclusion on the optimal way to adapt these pre-training methods for Wi-Fi signals, especially considering that natural language is discrete while signals are continuous. In the earliest works \cite{sun2021bert,guo2022wepos,wang2023radio}, researchers first applied rounding operations to the signals for tokenization and then directly input the tokens into language models. Our previous work \cite{zhao2024finding} highlighted the information loss problem associated with these methods and initiated a new paradigm that replaces the embedding layer of language models, enabling them to process continuous signal data directly. This was followed by subsequent works \cite{zhao2024mining,fan2024csi,gao2024multimodal,catak2025bert4mimo,yang2025wirelessgpt}. However, we also identified a significant challenge with this approach: the lazy loss function problem, where the model sometimes outputs the same value for all predictions. This value is often close to the mean of the entire dataset, leading to a low loss function value while the model fails to learn effectively. In our work, we employed a standardization layer and a discrimination loss to mitigate this issue. Yang et al. \cite{yang2024maskfi} proposed a novel method that combines two types of approaches in vision and Wi-Fi sensing. They used a Vision Transformer (ViT) \cite{dosovitskiy2020image} to process continuous Wi-Fi and image data while employing VQVAE \cite{van2017neural} to generate discrete ground truth for the MLM task. This method avoids information loss by using original data and addresses the lazy loss function problem by transforming the regression task into a classification task. Despite these advancements, there is still no consensus on which pre-training adaptation method for Wi-Fi data is optimal, indicating the need for further exploration.

\subsection{Future Directions}

For the future direction of multi-modal Wi-Fi sensing, in addition to the previously mentioned aspects such as exploring and developing cross-domain capacity, searching for optimal pre-training methods to make full use of limited datasets, developing high-efficiency dataset collection methods, and encouraging open source in the community, there are several other directions worth exploring. For example, LLM, LVM, LMM all demonstrate high generalization capacity attributed to scaling laws. Many domain-specific applications are beginning to construct foundation models transferred from these general models. Realizing that this approach is also useful in multi-modal Wi-Fi sensing can help us rapidly develop systems for new application scenarios with a minimal amount of fine-tuning data. Another potential direction is constructing a general model capable of processing heterogeneous data formats from different Wi-Fi receivers, which is important for developing a foundation model for Wi-Fi sensing, including multi-modal approaches. Additionally, multi-modal Wi-Fi sensing can be combined with various fields such as edge computing, federated learning, and embodied AI. For instance, some small robots \cite{bing2023lateral} have limited space, making it challenging to equip large devices like cameras. Currently, some control algorithms \cite{zhang2024autonomous} can only utilize IMU sensors. Given that many chips are equipped with Wi-Fi and Bluetooth receivers, it seems promising to combine data from these sources to enhance the robot's sensing capabilities.

\section{Conclusion}
Multi-modal Wi-Fi sensing represents a significant advancement over traditional Wi-Fi sensing by integrating complementary modalities such as vision and radar, thereby enhancing accuracy and robustness. This short survey has reviewed two primary approaches: multi-modal fusion at the input and feature levels, and enhanced training techniques like knowledge distillation and label generation. Despite its promise, the field faces challenges related to data scarcity, domain adaptation, and modality alignment. Future research should prioritize cross-domain generalization, efficient pre-training methods, and open-source collaboration to propel practical applications forward. As the field evolves alongside advancements in Wi-Fi 8 and multi-modal machine learning, multi-modal Wi-Fi sensing holds tremendous potential.

\bibliographystyle{ieeetr}
\bibliography{ref.bib}

% \section*{References}

% References follow the acknowledgments in the camera-ready paper. Use unnumbered first-level heading for
% the references. Any choice of citation style is acceptable as long as you are
% consistent. It is permissible to reduce the font size to \verb+small+ (9 point)
% when listing the references.
% Note that the Reference section does not count towards the page limit.
% \medskip

% {
% \small

% [1] Alexander, J.A.\ \& Mozer, M.C.\ (1995) Template-based algorithms for
% connectionist rule extraction. In G.\ Tesauro, D.S.\ Touretzky and T.K.\ Leen
% (eds.), {\it Advances in Neural Information Processing Systems 7},
% pp.\ 609--616. Cambridge, MA: MIT Press.

% [2] Bower, J.M.\ \& Beeman, D.\ (1995) {\it The Book of GENESIS: Exploring
%   Realistic Neural Models with the GEneral NEural SImulation System.}  New York:
% TELOS/Springer--Verlag.

% [3] Hasselmo, M.E., Schnell, E.\ \& Barkai, E.\ (1995) Dynamics of learning and
% recall at excitatory recurrent synapses and cholinergic modulation in rat
% hippocampal region CA3. {\it Journal of Neuroscience} {\bf 15}(7):5249-5262.
% }

\end{document}